Successive H-atom addition to solid OCS on compact amorphous solid water

Thanh Nguyen, Yasuhiro Oba, W. M. C. Sameera, Akira Kouchi, Naoki Watanabe

Institute of Low Temperature Science, Hokkaido University

N19W8, Kita-ku, Sapporo, Hokkaido 060-0819 Japan.

ABSTRACT

Carbonyl sulfide (OCS) is an abundant sulfur (S)-bearing species in the interstellar medium. It is present not only in the gas phase, but also on interstellar grains as a solid; therefore, OCS very likely undergoes physicochemical processes on icy surfaces at very low temperatures. The present study experimentally and computationally investigates the reaction of solid OCS with hydrogen (H) atoms on amorphous solid water at low temperatures. The results show that the addition of H to OCS proceeds via quantum tunneling, and further addition leads to the formation of carbon monoxide (CO), hydrogen sulfide ($H_2S$), formaldehyde ($H_2CO$), methanol ($CH_3OH$), and thioformic acid (HC(O)SH). These experimental results are explained by our quantum chemical calculations, which demonstrate that the initial addition of H to the S atom of OCS is the most predominant, leading to the formation of OCS-H radicals. Once the formed OCS-H radical is stabilized on ice, further addition of H to the S atom yields CO and $H_2S$, while that to the C atom yields HC(O)SH. We have also confirmed, in a separate experiment, the HC(O)SH formation by the HCO reactions with the SH radicals. The present results would have an important implication for the recent detection of HC(O)SH toward G+0.693–0.027.

1. INTRODUCTION

Sulfur (S)-bearing species have attracted great attention from astrochemical communities since the early 1970s when carbonyl monosulfide (CS) was detected as the first S-bearing species in the interstellar medium (ISM; Penzias et al. 1971). Since then, approximately 30 S-bearing species have been identified in the ISM to date, most of which have been detected only in the gas phase. Nevertheless, the total abundance of the observed gaseous S-bearing species is far from the cosmic abundance of sulfur, which is often denoted as the "sulfur depletion problem" (Ruffle et al. 1999; Vidal et al. 2017; Laas & Caselli 2019). The depletion of the S-bearing species in the gas phase implies that the "missing sulfur" may be present as a solid on interstellar grains. A S-bearing species often invoked for the source of the missing sulfur is hydrogen sulfide ($H_2S$; Codella et al. 2005). However, although $H_2S$ would mainly form on grains (Millar et al. 1986; Millar & Herbst 1990), it has never been positively identified as a solid in the ISM probably because of its desorption from grains into the gas phase upon formation via chemical desorption and transformation into other S-bearing species via the energetic processes induced by ultraviolet photons and cosmic rays, as evidenced by recent laboratory studies (Oba et al. 2018, 2019; Ferrante et al. 2008; Garozzo et al. 2010; Jimenez-Escobar & Muñoz Caro 2011; Jimenez-Escobar et al. 2012, 2014).

One of the other candidates for the missing sulfur might be carbonyl sulfide (OCS; Codella et al. 2005), which has been confirmed both in the gas and solid phases of the ISM with abundances of $\sim 1.6 \times 10^{-9}$ (in average) vs. $H_2$ (Goldsmith & Linke 1981) and $4 \times 10^{-4}$ vs. $H_2O$ in W33A (Palumbo et al. 1995), respectively. Laboratory experiments have demonstrated that OCS can be produced by the photochemical reactions of $H_2S$-containing ices under the ISM conditions (Ferrante et al. 2008; Garozzo et al. 2010; Jimenez-Escobar & Muñoz Caro 2011; Jimenez-Escobar et al. 2012, 2014). In addition, the OCS formation on the interstellar grains via nonenergetic processes, such as chemical processes that are not induced by ultraviolet photons and cosmic rays, has been investigated computationally (Adriaens et al. 2010). The following reactions are proposed to occur, even at low temperatures:

$$CO + S \rightarrow OCS, \tag{1}$$

$$CS + O \rightarrow OCS, \tag{2}$$

$$CO + HS \rightarrow H\text{-}SCO \rightarrow OCS + H, \tag{3}$$

$$CS + OH \rightarrow H\text{-}OCS \rightarrow OCS + H. \tag{4}$$

In contrast to the OCS formation, much less attention has so far been paid to its reactivity on grains. In particular, no experimental studies have yet been conducted on the nonenergetic reaction of OCS under the ISM conditions. Given that OCS is present as a solid on interstellar grains, it very likely reacts with other species, such as H atoms and OH radicals, which play important roles for the molecular evolution on grains (Watanabe & Kouchi 2008; Hama & Watanabe 2013; Tsuge & Watanabe 2021). When OCS reacts with the H atoms on the dust grain surface, we can find the following three reaction channels depending on which atom is attacked by a H atom:

$$OCS + H \rightarrow H\text{-}OCS \text{ (addition to O atom)}, \tag{5}$$

$$\rightarrow OC(H)S \text{ (addition to C atom)}, \tag{6}$$

$$\rightarrow OCS\text{-}H \text{ (addition to S atom)}. \tag{7}$$

Each reaction had a large barrier (>~25 kJ mol$^{-1}$; Rice et al. 1998; Saheb et al. 2012). Although these reactions were investigated in the gas phase both experimentally (Lee et al. 1977) and computationally (Rice et al. 1998; Saheb et al. 2012), it was not at all on low-temperature surfaces, such as amorphous solid water (ASW), which are more relevant to interstellar grain conditions. ASW has some catalytic properties that enhance the chemical reactions on the surface (Hama & Watanabe 2013); thus, the reaction OCS + H may yield some products that differ from what are expected in the gas phase. Note that if an OCS-H radical, which may form in Reaction (7), further reacts with an additional H atom, thioformic acid (HC(O)SH) may form as follows:

$$OCS\text{-}H + H \rightarrow HC(O)SH. \tag{8}$$

Very recently, Rodríguez-Almeida et al. (2021) detected the transconfiguration of this molecule toward the galactic center quiescent cloud G+0.693–0.027 with an abundance of ~1 × 10$^{-10}$ vs. H$_2$. They also proposed that the sequential hydrogenation to OCS is one of the possible pathways toward HC(O)SH formation.

The present paper presents experimental and computational results on the surface reactions of OCS with H atoms on ASW to better understand the sulfur chemistry in dense molecular clouds.

## 2. EXPERIMENTS

All experiments were performed using the Apparatus for Surface Reaction in Astrophysics (ASURA) system. The details of the ASURA system are described elsewhere (Watanabe et al. 2006; Nagaoka et al. 2007; Nguyen et al. 2020). The ASURA system consists of a stainless-steel reaction chamber with a base pressure of 10$^{-8}$ Pa, an Al reaction substrate coated with a thin layer of Mg$_2$SiO$_4$ attached to a He cryostat, an atomic source, a quadrupole mass spectrometer (QMS), and a Fourier-transform infrared spectrometer (FTIR).

The reactions of the solid OCS with the H atoms were studied on compact amorphous solid water (c-ASW) made by vapor deposition of H$_2$O onto the reaction substrate maintained at 110 K. The c-ASW layer thickness was adjusted to ~30 monolayers (MLs; 1 ML = 1 × 10$^{15}$ molecules cm$^{-2}$). The substrate was cooled to 10 K after the c-ASW layer production. Gaseous OCS (99.9% in purity, Taiyo Nippon Sanso Corp.) was vapor-deposited onto the predeposited c-ASW with a deposition rate of 1 ML minute$^{-1}$. The OCS layer thickness was adjusted to 1 ML, which was estimated using the peak area of the CO stretching band at 2049 cm$^{-1}$ and its absorption coefficient of 1.5 × 10$^{-16}$ cm molecule$^{-1}$ (Hudgins et al. 1993). The predeposited OCS layer on c-ASW was exposed to the H (or D) atoms produced through the dissociation of H$_2$ (D$_2$) in a microwave-discharged plasma in an atomic chamber. The formed H (D) atoms were cooled to 100 K by multiple collisions with the inner wall of the Al pipe at 100 K. Following the method of Oba et al. (2014), the flux of the H and D atoms was estimated as 1.5 × 10$^{14}$ and 4.6 × 10$^{14}$ atoms cm$^{-2}$ s$^{-1}$, respectively.

In a separate experiment, gaseous OCS was co-deposited with the H atoms onto a pre-deposited c-ASW layer (30 ML) to gain sufficient yields for the FTIR detection. The deposition rate of the gaseous OCS was ~$2.6 \times 10^{-2}$ ML min$^{-1}$; thus, the OCS/H mixing ratio was $2.9 \times 10^{-3}$. The reaction products were monitored in-situ by the FTIR with a 2 cm$^{-1}$ resolution. The reactants and the products desorbed from the substrate were monitored using the QMS via the temperature-programmed desorption (TPD) method with a 4 K min$^{-1}$ ramping rate.

## 3. COMPUTATIONAL METHODS

The reaction mechanism of OCS and H on ice was calculated using the computational methods. An ice cluster model consisting of 76 water molecules (Figure A1) was prepared from the two-dimensional periodic ASW ice model in previous studies (Andersson et al. 2006; Nguyen et al. 2021). The positions of the outermost water molecules kept frozen as in the structural models by Andersson et al. (2006). Geometry optimizations were performed using the wB97X-D functional (Chai & Head-Gordon 2008), as implemented in the Gaussian 16 program (version C.01; Frisch et al. 2016). The outermost water molecules in the side and bottom walls of the water cluster were frozen to avoid structural deformations and save computational costs. The def2-TZVP (Weigend & Ahlrichs 2005) basis sets were employed for all water molecules. The vibrational frequency calculations confirmed that the optimized structures were local minimum (i.e., no imaginary frequency) or transition states (i.e., one imaginary frequency), and determined the zero-point energy. The connectivity between the local minimum and the transition states was confirmed by performing pseudo-intrinsic reaction coordinate calculations, where 20 steps for the forward and backward directions were followed starting from the transition-state geometry.

In the reactions between two radicals, the spin of the two radicals can be ferromagnetically or antiferromagnetically coupled at a short-range separation. Therefore, the wave function of the molecular system is multiconfigurational and cannot be treated by DFT. To overcome this issue, we used our own

(two-layer) N-layer integrated molecular orbital molecular mechanics (ONIOM; Chung et al. 2015; Sameera & Maseras 2018) method for structure optimizations. The coupled-cluster singles and doubles (CCSD; Čížek 1969; Purvis & Bartlett 1982; Watts et al. 1993) method and cc-pVTZ (Dunning 1989; Kendall et al. 1992; Woon & Dunning 1993) basis sets were used to treat the two radicals. The ice cluster was included in the ONIOM low layer and computed using the wB97X-D functional and the def2-TZVP basis sets. Potential energy of the optimized structures were calculated as the single-point energy using the ONIOM(CCSD(t)/aug-cc-pVTZ:wB97X-D/def2-TZVP) calculations.

## 4. EXPERIMENTAL RESULTS

### 4.1. Reaction of the Predeposited OCS with the H or D atoms

Figure 1(a) shows the FTIR spectrum of the solid OCS deposited on c-ASW at 10 K. The observed peak patterns were well consistent with those in the previous studies, where the CO stretching band at 2049 cm$^{-1}$ was predominant (Hudgins et al. 1993; Ferrante et al. 2008). Figure 1(b) depicts the variation in the difference spectra after exposure to the H atoms for up to 1 hr. The CO stretching band intensity decreased with the atom exposure times, showing the OCS depletion. In addition, a small peak newly appeared at ~1040 cm$^{-1}$, which was expected to be derived from the CO stretching band of the solid CH$_3$OH, which would be formed by sequential hydrogenation to the solid CO on c-ASW:

$$CO + H \rightarrow HCO, \qquad (9)$$

$$HCO + H \rightarrow H_2CO, \qquad (10)$$

$$H_2CO + H \rightarrow CH_3O\ (H_2COH), \qquad (11)$$

$$CH_3O\ (H_2COH) + H \rightarrow CH_3OH, \qquad (12)$$

where reactions (9) and (11) proceed via quantum tunneling (Watanabe & Kouchi 2002; Watanabe et al. 2003). The peaks derived from formaldehyde (H$_2$CO) were obscured on the spectra probably because of its conversion to CH$_3$OH through reactions (11) and (12). According to our quantum chemical calculations

(Section 5), CO can be formed as follows by the OCS-H radical hydrogenation, which is the product of the OCS hydrogenation by Reaction (7):

$$\text{OCS-H} + \text{H} \rightarrow \text{H}_2\text{S} + \text{CO}. \quad (13)$$

The CO and $H_2S$ formation is thermodynamically more favored over the HC(O)SH formation by Reaction (8); thus, the formation of the CO-bearing species, such as $CH_3OH$, is reasonable under the present experimental conditions (see Section 5 for details). In contrast, the solid $H_2S$ was not clearly identified from the IR spectrum (Figure 1(b)) because of the efficient loss from the ice surface by chemical desorption through these reactions (Oba et al. 2018, 2019):

$$\text{H}_2\text{S} + \text{H} \rightarrow \text{HS} + \text{H}_2, \quad (14)$$

$$\text{HS} + \text{H} \rightarrow \text{H}_2\text{S}. \quad (15)$$

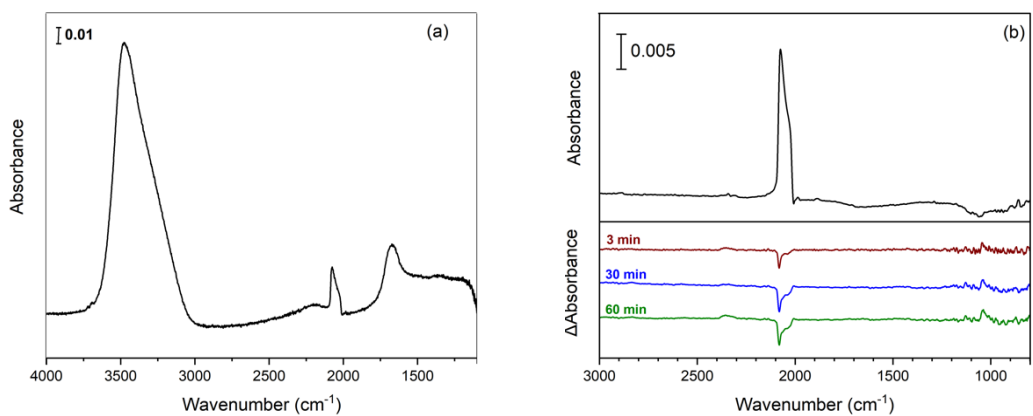

Figure 1: (a) Fourier-transform infrared profile of OCS on c-ASW (30 ML) at 10 K. (b) Variation in the difference spectra of the solid OCS after exposure to H atoms for 3, 30, and 60 minutes at 10 K. The initial C–O stretching band is shown as a reference. The $\nu_8$ band of $CH_3OH$ was observed at 1040 cm$^{-1}$ in the difference spectra while $H_2S$ was not at 2550 cm$^{-1}$.

Note that the formation of CO and $H_2S$ was clearly confirmed in the codeposition (OCS and H atoms) experiments shown in the next section. Although the OCS decrease was also observed after exposure to the

D atoms, the product peak identification was not conclusive probably due to the limited amounts of OCS, which reacted with the D atoms.

Figure 2 shows the variations in the relative abundance of the OCS after exposure to the H or D atoms for up to 2 hr. The OCS decrease was approximately 16% and 8% of the initial abundance after reactions with the H and D atoms, respectively, for 2 hr. The column density of the $CH_3OH$ formed after exposure to the H atoms for 2 hr was approximately 0.15 ML, which was equivalent to ~90% of the OCS loss. The observed predominance of $CH_3OH$ in the products was well consistent with the theoretical prediction given in the next section.

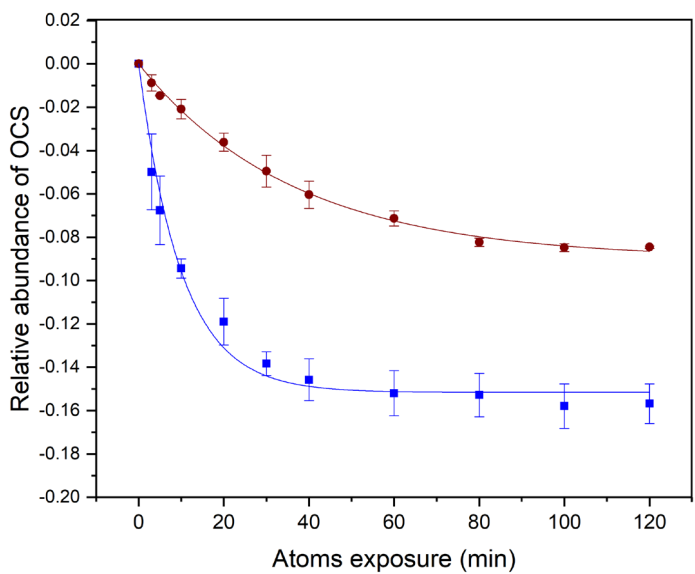

Figure 2: Variation in the peak areas of the C–O stretching band for the solid OCS as a function of exposure to H (blue squares) and D (red circles) atoms up to 2 hr on c-ASW at 10 K.

Assuming that OCS is reduced by reactions with H or D atoms only, the observed variations in the relative abundance of OCS can be described using the following rate equation:

$$d[OCS]/dt = -k_X[X][OCS], \qquad (16)$$

where [OCS] and [X] are the surface number density of the OCS and X atoms (X = H or D) on ices,

respectively, and $k_X$ is the rate coefficient of the X addition reactions. We obtained the following by integrating Eq. (16):

$$\Delta[OCS]_t/[OCS]_0 = \alpha(1 - \exp(-k'_X t)), \qquad (17)$$

where $\Delta[OCS]_t$ and $[OCS]_0$ represent the variation in the abundance of the solid OCS at time $t$ and the initial OCS abundance, respectively, and $k'_X$ ($= k_X[X]$) is the effective rate constant for the H or D addition reactions. By fitting the plots in Figure 2 into Eq. (17), we obtained the $k'_H$ and $k'_D$ values as $(1.7 \pm 0.2) \times 10^{-3}$ and $(4.6 \pm 0.2) \times 10^{-4}$ s$^{-1}$, respectively. According to Kuwahata et al. (2015), the number density ratio of the H and D atoms ([D]/[H]) on porous amorphous and crystalline H$_2$O ices was ~4 and ~9, respectively, when the flux of atoms was on the order of $10^{14}$ atoms cm$^{-2}$ s$^{-1}$. Assuming that the [D]/[H] on c-ASW is a value between them, the ratio of $k_H/k_D$ is roughly estimated from the $k'_H/k'_D$ value (3.6) and the following relationship:

$$k_H/k_D = k'_H/k'_D \times [D]/[H]. \qquad (18)$$

Under the assumption above, the $k_H/k_D$ value ranges from 14 to 32, strongly suggesting that the H and D additions to the solid OCS on c-ASW proceeded via quantum tunneling.

4.2. Identification of products after Codeposition of OCS with the H atoms

Figure 3 shows the FTIR spectrum of the solid OCS codeposited with the H atoms for 4 hr at 10 K and that of the pure solid OCS for comparison. Unlike the OCS predeposition experiment shown above, various peaks appeared after the codeposition of OCS with the H atoms, strongly indicating that the reactions with the H atoms actually proceeded to result in the formation of various products. Table 1 summarizes the wavenumbers of the observed peaks. A sharp peak that appeared at 2136 cm$^{-1}$ was reasonably assigned to the solid CO (Gerakines et al. 1995). This assignment was also supported by the disappearance of this peak when the substrate temperature was raised to 100 K (Fig. 4). According to the quantum chemical calculations in the gas phase, CO may be formed through the spontaneous dissociation of the OCS-H radical after Reaction (7) (Rice et al. 1998; Saheb et al. 2012). However, in this study, the formed OCS-H radicals were clearly stabilized by the interactions with water ices, which prevented their dissociation into CO and SH (Section 5). Thus, the formed OCS-H radical should further react with the H atoms without a barrier, resulting in the formation of both CO and $H_2S$ via Reaction (13). The $H_2S$ formation was strongly inferred from the appearance of a broad peak at 2550 cm$^{-1}$, which was well consistent with the SH stretching band of the solid $H_2S$ (Oba et al. 2018). In addition, several peaks derived from $H_2CO$ at 1500, 1720, and ~2800 cm$^{-1}$ (Schutte et al. 1993; Hidaka et al. 2004) and $CH_3OH$ at 1030, ~1450, and ~2830–3055 cm$^{-1}$ (Nagaoka et al. 2007) were observed. The OH stretching band of $CH_3OH$ was not clearly identified at 3000–3500 cm$^{-1}$ due to overlapping with the OH stretching band of c-ASW. These two molecules were reasonably formed by sequential hydrogenation to solid CO, as mentioned earlier.

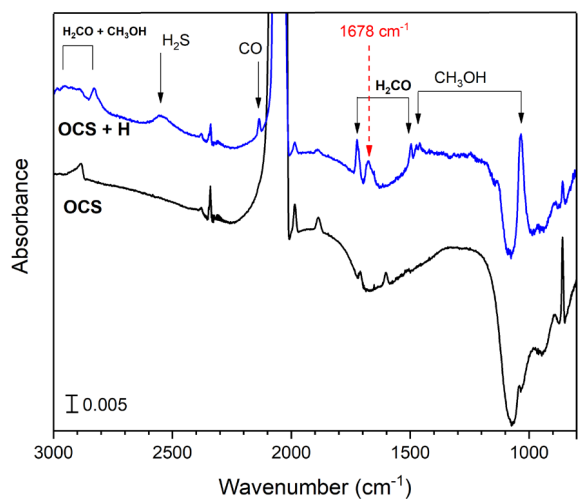

Figure 3: FTIR spectra of the pure OCS (black) and the codeposition OCS and H atoms (blue) on c-ASW at 10 K. The observed bump at ~1050 cm$^{-1}$ could be caused by the interaction of $H_2O$ with silicates (Potatov et al. 2018; Nguyen et al. 2020).

The spectral profile varied when a large part of the unreacted OCS desorbed from the substrate above 100 K (Figure 4). For example, CO and a large part of the H$_2$CO and H$_2$S features disappeared due to their desorption at temperatures below 110 K. In contrast, multiple peaks appeared at 1200–1500 cm$^{-1}$ with an increasing temperature. The increase in the peak intensities upon warming up may be caused by the aggregation of molecules, which were distributed in the solid OCS at 10 K, to form a solid (Hidaka et al. 2011). In addition, the new species were also likely formed at higher temperatures by reactions with reactive species, which cannot diffuse inside the bulk OCS solid at 10 K.

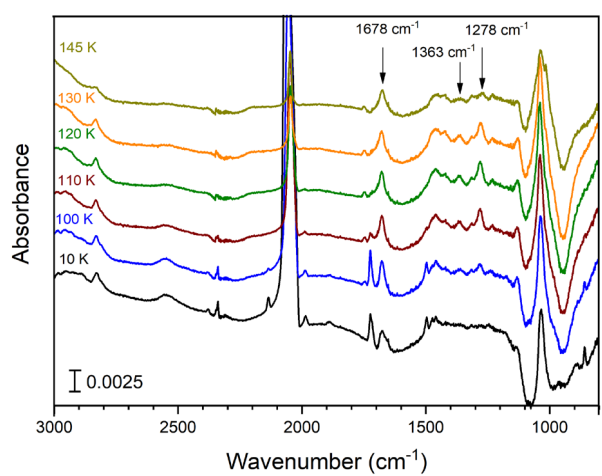

Figure 4: Variation in the different spectra after the codeposition of OCS and H atoms on c-ASW when the surface temperature increases from 10 to 145 K. The observed bump at ~1050 cm$^{-1}$ could be caused by the interaction of H$_2$O with silicates (Potatov et al. 2018; Nguyen et al. 2020).

Aside from the new peaks that appeared at 1200–1500 cm$^{-1}$, the relatively strong peak at 1678 cm$^{-1}$ (Figures 3 and 4), 1363 cm$^{-1}$, and 1278 cm$^{-1}$, as well as a small peak at 1749 cm$^{-1}$ (Figure 4) cannot be assigned to any of the abovementioned species. As shown in the next section, our computational calculations proposed that following the H$_2$S and CO formation (Reaction (13)), the HC(O)SH formation is possible in the present experiment (Reaction (8)). By comparing the observed absorption bands with those of HC(O)SH in the liquid state (Gattow & Engler 1971) and in solid rare gases (Lignell et al. 2021), a strong peak at 1678 cm$^{-1}$ might be derived from its CO stretching mode (Table 1). A weak absorption at 2550 cm$^{-1}$ observed at >100 K (Figure 4), which is typically assigned to the SH stretching mode, also implied the presence of HC(O)SH on the substrate. The CH stretching bands for HC(O)SH should appear at ~2850 cm$^{-1}$ (Fausto et al. 1989; Lignell et al. 2021), but cannot be distinguished from those of CH$_3$OH (Nagaoka et al. 2007). Although the solid HC(O)SH also had an absorption of the CH bending at around 1350 cm$^{-1}$ (Lignell et al. 2021), we expect that the observed peak at 1363 cm$^{-1}$ may be derived from other species due to the following reasons: first, this peak almost disappeared at ~145 K, while the peak at 1678 cm$^{-1}$ was still observed (Figure 4); second, the peak intensity of the CH bending band was predicted to be approximately 3% of the CO stretching band in computational calculations (Lignell et al. 2021), which is clearly different from the present result (~30%; Figure 4). Unfortunately, the presence of HC(O)SH was not confirmed by the TPD-QMS experiment, possibly because of the small amount in the product. Nevertheless, we are confident that the HC(O)SH formation would be the most reasonable explanation for the observed results under the present experimental conditions.

The 1278 cm$^{-1}$ peak cannot be assigned in this study. One of the possible candidates for the 1278 cm$^{-1}$ peak is thionformic acid (HC(S)OH), which is the structural isomer of thioformic acid (1265 cm$^{-1}$ in solid argon; Lignell et al. 2021). HC(S)OH may be formed by hydrogenation to the C atom of the H-OCS radicals; however, the H-OCS formation was not favored due to the large barrier of 0.97 eV (Section 5). Another possible channel for the HC(S)OH formation is hydrogenation to the O atom of the OC(H)S radicals. However, the OC(H)S formation by Reaction (6) was not favored on ice (Section 5); hence, the HC(S)OH formation was difficult by these two processes. Another possible product after sequential hydrogenation to the solid OCS is HOCSH, which has a linear structure with a carbon–sulfur triple bond. HOCSH could have a contribution to the absorption at 1278 cm$^{-1}$ because a molecule that possesses a CS triple bond typically has absorption in the same region (Schreiner et al. 2009; Denis & Iribarne 2011).

Further confirmation is necessary for their assignments.

We next focus on the formation mechanism of HC(O)SH under the present experimental conditions. As shown in the quantum chemical calculations, the HC(O)SH formation via sequential hydrogenation to OCS (reactions (7) and (8)) probably occurs under the present experimental conditions. In addition, Rodríguez-Almeida et al. (2021) expected that the HCO reaction with the SH would result in the HC(O)SH formation:

HCO + SH → HC(O)SH.                              (19)

Note that the HCO and SH radicals would be formed by reactions (9) and (14), respectively. Hence, if Reaction (19) can proceed, there are at least two channels toward the HC(O)SH formation under the present experimental conditions.

To test this hypothesis, we performed an additional experiment for the reaction of HCO and SH, which were produced by reactions (9) and (14), respectively, on c-ASW. Figure 5 shows the FTIR spectrum obtained after the codeposition of three components (i.e., CO, $H_2S$, and

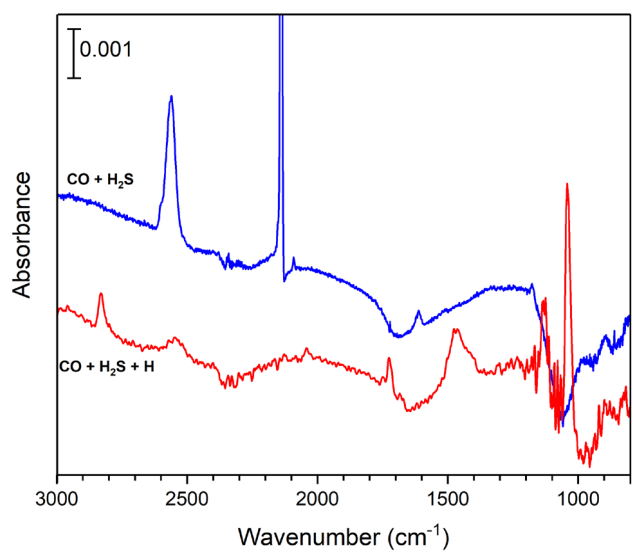

Figure 5: Comparison of the FTIR spectra between two codeposition experiments on c-ASW at 10 K: top for CO + $H_2S$ (blue) and bottom for CO + $H_2S$ + H (red). Cartesian coordinates of the optimized structures

can be found in the tar.gz package of supporting information.

H atoms) onto c-ASW at 10 K and after the codeposition of two components (i.e., CO and $H_2S$) for comparison. The ratio of CO to $H_2S$ was adjusted to 1:1, which was equal to the OCS hydrogenation experiment. In the three-component experiment, both CO and $H_2S$ would be largely consumed by the reactions with the H atoms, resulting in the formation of $H_2CO$ (~1500, 1726 and ~2900 cm$^{-1}$) and $CH_3OH$ (1040, 1135, ~1460 and ~2900 cm$^{-1}$), which were formed by sequential hydrogenation to CO (reactions (9)–(12)). In addition, one small peak appeared at 1678 cm$^{-1}$, as in the case of the OCS + H experiments (Figure 3). Based on the band position and the experimental conditions, where both the HCO and HS radicals are available, we strongly expected this peak to be attributable to the HC(O)SH formed by reaction (19). Another small peak that appeared at 2041 cm$^{-1}$ might be derived from the CO stretching band of OCS. Several reactions could yield OCS using CO and HS (Adriaens et al. 2010); however, elucidating its formation pathways is beyond the scope of the present study.

The S-bearing species, including HC(O)SH and OCS, are apparently less than the CO-bearing species (e.g., $H_2CO$ and $CH_3OH$) in the three-component codeposition products. If both HCO and SH radicals are present nearby ices, they may recombine to form HC(O)SH without a barrier. In contrast, both radicals would not easily diffuse on ice at 10 K; thus, they should preferentially react with the H atoms, which can easily diffuse on ice surfaces, even at 10 K (Hama & Watanabe 2013), resulting in the formation of $H_2CO$, $CH_3OH$, and $H_2S$. The formed $H_2S$ desorbs from ices upon formation by chemical desorption (Oba et al. 2018), which would have resulted in a significant loss of $H_2S$ compared to the two-component (CO and $H_2S$) codeposition experiment (Figure 5).

5. Computational results

First, we calculated the binding energies between OCS and ASW. Figure A2 depicts the computed binding energies. A range of binding energies, i.e., 0.12–0.21 eV (1393-2437 K), was observed, where the computed average binding energy of 0.16 eV (1857 K) indicated that OCS binding on ice is possible. We used one of the OCS-bound ice structures to calculate the reaction mechanism between OCS and H on ASW. Figure 6 presents the calculated potential energy surfaces of the reactions between OCS and H in ASW.

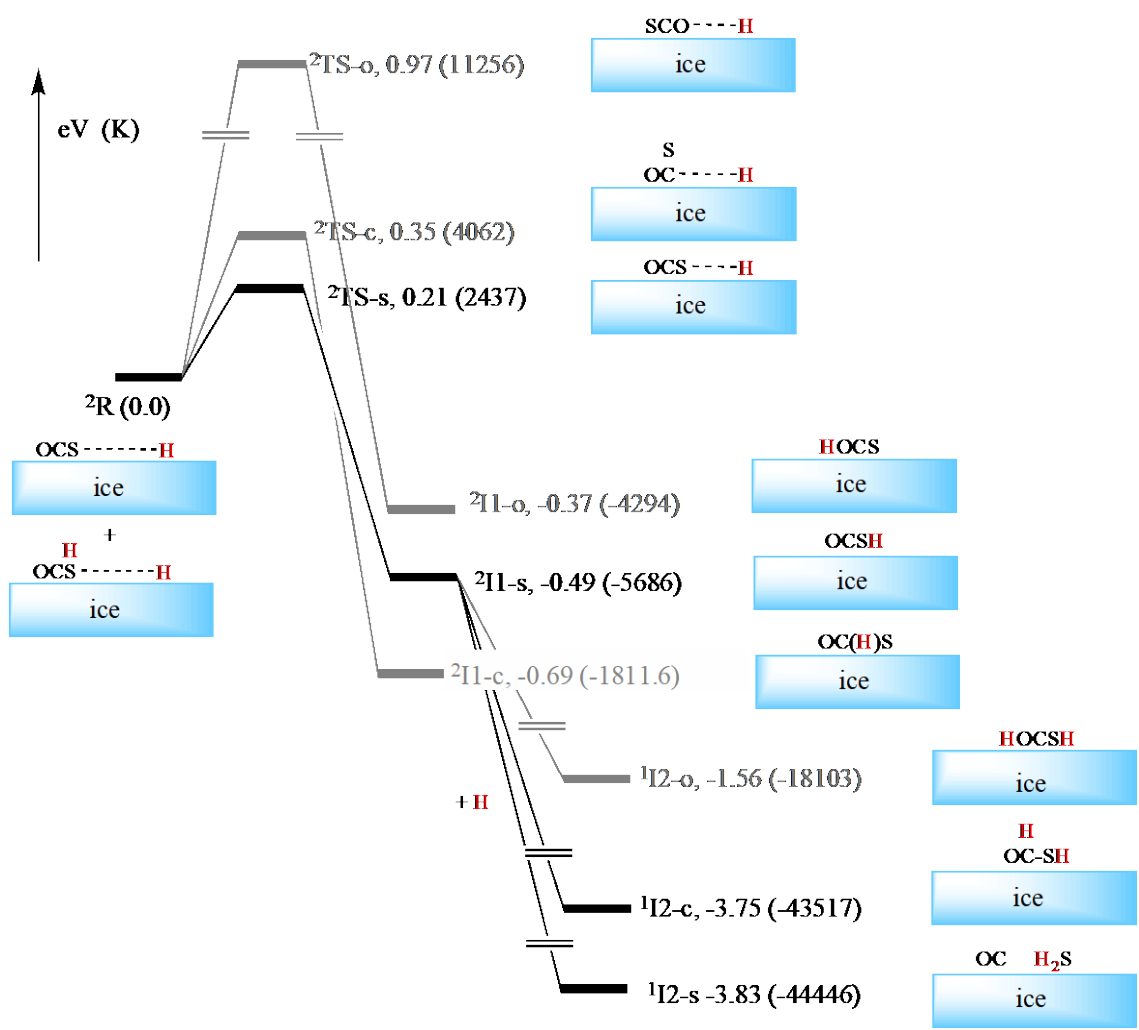

Figure 6. Computed potential energy surfaces for the reaction between OCS and H on ice from wB97X-D/def2-TZVP structure optimizations of the full system. Potential energy sum of OCS-H (doublet state) on ice and OCS(H)-H (open-shell single state) on ice were taken as the reference energy. Lines in black show the main reaction paths. The multiplicity of the local minima or transition states is shown in superscript.

Geometries of the optimized transition states are shown in Figure 7. The calculated energies of the optimized stationary points and imaginary frequencies are shown in Table A1.

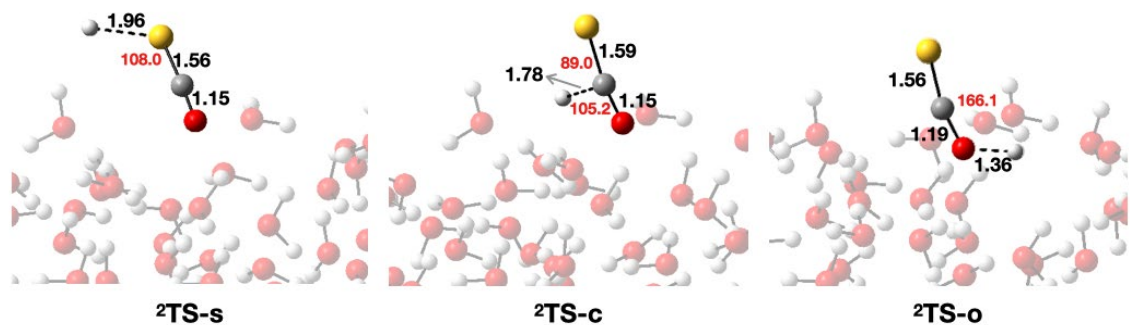

Figure 7. Geometries of the optimized transition states, ²TS-s, ²TS-c, and ²TS-o.

H atom attack at the S atom of OCS showed the lowest energy barrier (TS-s, 0.21 eV or 2437 K), which gave rise to OCS-H on ice. The computed reaction barriers for the H atom attack at the C and O atoms of OCS were 0.35 eV (or 4062 K, ²TS-c) and 0.97 eV (or 11256 K, ²TS-o) higher than $TS_C$, leading to OC(H)S and H-OCS on ice. The computed reaction path ratio for the product formation calculated using the Boltzmann distribution of the transition states was OCS-H:OC(H)S:H-OCS = 100:0:0. Therefore, the reaction between OCS and H atoms on ice would produce OCS-H. Even though the resulting OC(H)S is thermodynamically more stable than OCS-H, its concentration would be negligible.

The computed binding energies of the OCS-H radical on ice were in the range of 0.19–0.46 eV (2205-5338 K; Figure A3). The computed average binding energy was 0.31 eV, which is stronger than that of OCS on ice (i.e., 0.16 eV). Dipole moment of OCSH (1.68 D) is significantly larger than OCS (0.62 D). Therefore, C–O and S–H groups of OCSH can strongly interact with the dinging-H and dangling-O on ice, leading to strong binding on ice. Thus, the OCS-H bound on ice would react with H. Accordingly, we introduced a H atom to the molecular system, and possible reaction mechanisms between the H atom the OCS-H radical on ASW were explored. In principle, three reaction paths, (i) OCS-H + H → $H_2S$ + CO, (ii)

OCS-H + H → HC(O)SH, and (iii) OCS-H + H → HOCSH, are possible. In reaction (i), H atom attacked the S atom of OCS, giving rise to $H_2S$ and CO, which are 3.35 eV more stable than OCS-H. If H attacked the C atom of OCS (i.e., reaction (ii)), HC(O)SH can be formed, and is 3.26 eV more stable than OCS-H. This finding is in agreement with our experimental observations. According to computed relaxed potential energy surfaces, formation of The $H_2S$ + CO and HC(O)SH products barrierless processes (Figure A4). In the case of reaction (iii), H attacked the O atom of OCS, leading to HOCSH. However, the relaxed potential energy surface for this reaction path indicates a barrier (Figure A4), and therefore formation of HOCSH on ice may be difficult.

We have also calculated potential energy surfaces for the reactions between OCS and H in gas phase (Figure A5). In the absence of the ice surface, H atom attack at the S atom of OCS is still the lowest energy path with a barrier of 0.17 eV (1973 K, $^2$TS-s′), where the reaction barrier is slightly smaller compared to the analogous reaction on ice. Subsequent hydrogenation of the OCS-H is highly exothermic compared to the analogous processes on ice, giving rise to CO, $H_2S$, or HC(O)SH.

## 6. Astrochemical implications and conclusions

Rodríguez-Almeida et al. (2021) compared the observed abundances of S-bearing species (i.e., CS, $H_2CS$, $CH_3SH$, $C_2H_5SH$, and HC(O)SH) with their O-bearing counterparts (i.e., CO, $H_2CO$, $CH_3OH$, $C_2H_5OH$, and HC(O)OH) toward G+0.693. The observed similar trend between the S- and O-bearing species implied similar chemistry for the formation of both chemical groups. The O-bearing counterpart of HC(O)SH, which is HC(O)OH (formic acid), would not be formed on ices through sequential addition of H to $CO_2$ (the O-bearing counterpart of OCS) due to the very high activation barrier (~1.1 eV; Yu et al. 2001); hence, it is often considered that HC(O)OH is formed by the energetic processes induced by the ultraviolet photons and cosmic rays in interstellar ices (Hudson & Moore 1999; Watanabe et al. 2007; Bennett et al. 2011). In addition, the HC(O)OH formation by nonenergetic processes, such as the addition of H to a HOCO radical, was also proposed (Ioppolo et al. 2011), which may be more relevant to the HC(O)SH formation demonstrated in the present study. Organo-sulfur species and pure sulfur allotropes were predicted as the main sources of the missing sulfur on grains (Laas & Caselli 2019; Shingledecker et al. 2020); thus, further experimental studies on the formation of these species are highly necessary for clarifying the sulfur depletion problem.

For the first time, we performed experimental studies on the hydrogenation of solid OCS on c-ASW at 10 K. The reaction cannot proceed thermally at 10 K because of the presence of a large activation barrier for the H-addition reaction of 0.21 eV (= 2436 K). We demonstrated that the reaction would actually proceed through quantum tunneling, as evident in the large isotope effect. The quantum chemical calculations predicted that the main product by the sequential hydrogenation to the solid OCS is CO and $H_2S$, which is consistent with the experimental results. The formed CO and $H_2S$ further reacted with the H atoms, resulting in the formation of CO-bearing species, such as $H_2CO$ and $CH_3OH$, as well as the loss of $H_2S$ from the ice surfaces by chemical desorption. We tentatively, but with strong confidence, assigned thioformic acid (HC(O)SH) on the IR spectrum of the product. HC(O)SH may be formed either or both by sequential hydrogenation to OCS or the reaction HCO and SH radicals formed from CO and $H_2S$ after reactions with H atoms, respectively. A recent astronomical observation toward G+0.693–0.027 predicted the HC(O)SH formation by surface reactions on interstellar grains (Rodríguez-Almeida et al. 2021), which is well consistent with the present result. In addition to HC(O)SH, we observed a couple of peaks in the product; however, these peaks were not assigned to any species considered in the present study. Further studies on the assignment of these new species are required, and these products may likely constitute a part of the missing sulfur in dense molecular clouds.

Table 1: Infrared absorption bands after co-deposition of OCS + H and CO + H$_2$S + H on c-ASW.

| Band position (cm$^{-1}$) | | | Assignment | References |
|---|---|---|---|---|
| OCS + H | H$_2$S + CO + H | Calculation | | |
| 860 | - | | OCS (C=S str) | Ferrante et al. (2008) |
| 1030 | 1040 | | CH$_3$OH (C=O str) | Nagaoka et al. (2007) |
| 1131 | 1135 | | CH$_3$OH (C–H rock) | Nagaoka et al. (2007) |
| 1278 | - | | HOCSH? (C≡S str?) | Denis & Iribarne (2011) |
| 1363 | - | | unknown | |
| ~1450 | ~1460 | | CH$_3$OH (C–H bend) | Nagaoka et al. (2007) |
| 1500 | ~1500 | | H$_2$CO (C–H bend) | Schutte et al. (1993) |
| 1678 | 1678 | 1769 | HC(O)SH (C=O str) | Gattow and Engler (1971); Lignell et al. (2021) |
| 1710 | - | | OCS (C=O str) | Ferrante et al. (2008) |
| 1720 | 1726 | | H$_2$CO (C=O bend) | Schutte et al. (1993) |
| 1749 | 1749 | | unknown | |
| 2049 | 2041 | | OCS (C=O str) | Ferrante et al. (2008) |
| 2136 | 2130 | | CO (C=O str) | Gerakines et al. (1995) |
| 2550 | 2550 | 2685 | H$_2$S (S-H str); HC(O)SH (S-H str) | Fathe et al. (2006); Lignell et al. (2021) |
| 2830–3055 | 2830–2987 | 3033 | H$_2$CO (C–H str), CH$_3$OH (C–H str); HC(O)SH (C–H str) | Schutte et al. (1993); Nagaoka et al. (2007); Lignell et al. (2021) |

Note. Our calculation results for HC(O)SH are also shown for comparison.

Acknowledgements

This work is partly supported by JSPS KAKENHI grant Nos. JP17H06087, JP17H04862, JP19K03940, JP19K23449, JP21H04501, and JP21K13974. The computations were performed using Research Center for Computational Science, Okazaki, Japan.

Appendix

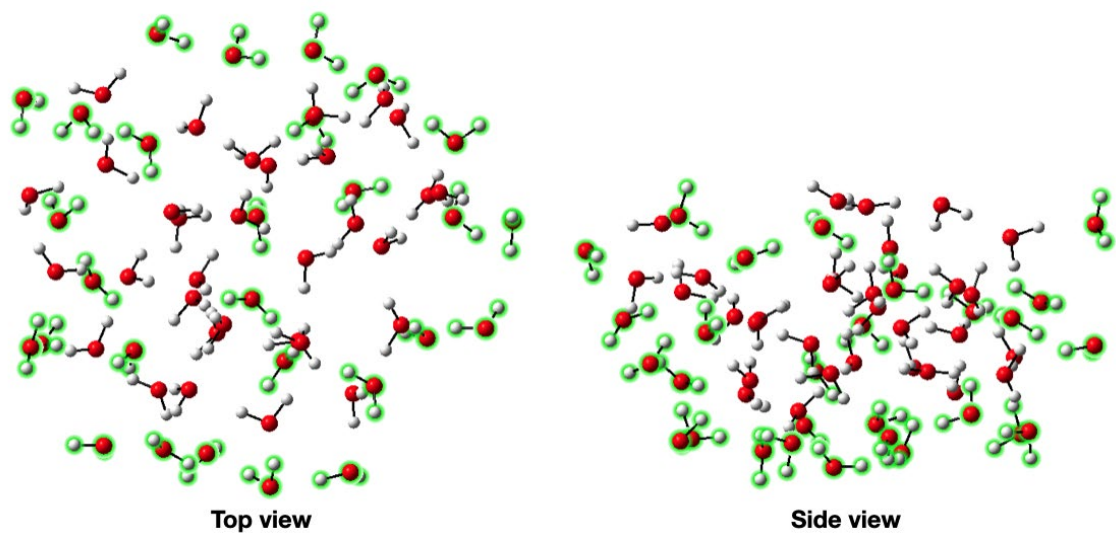

Figure A1. Ice cluster model consisting of 76 $H_2O$ molecules to replicate ASW. Water molecules highlighted in green were frozen during the structure optimizations.

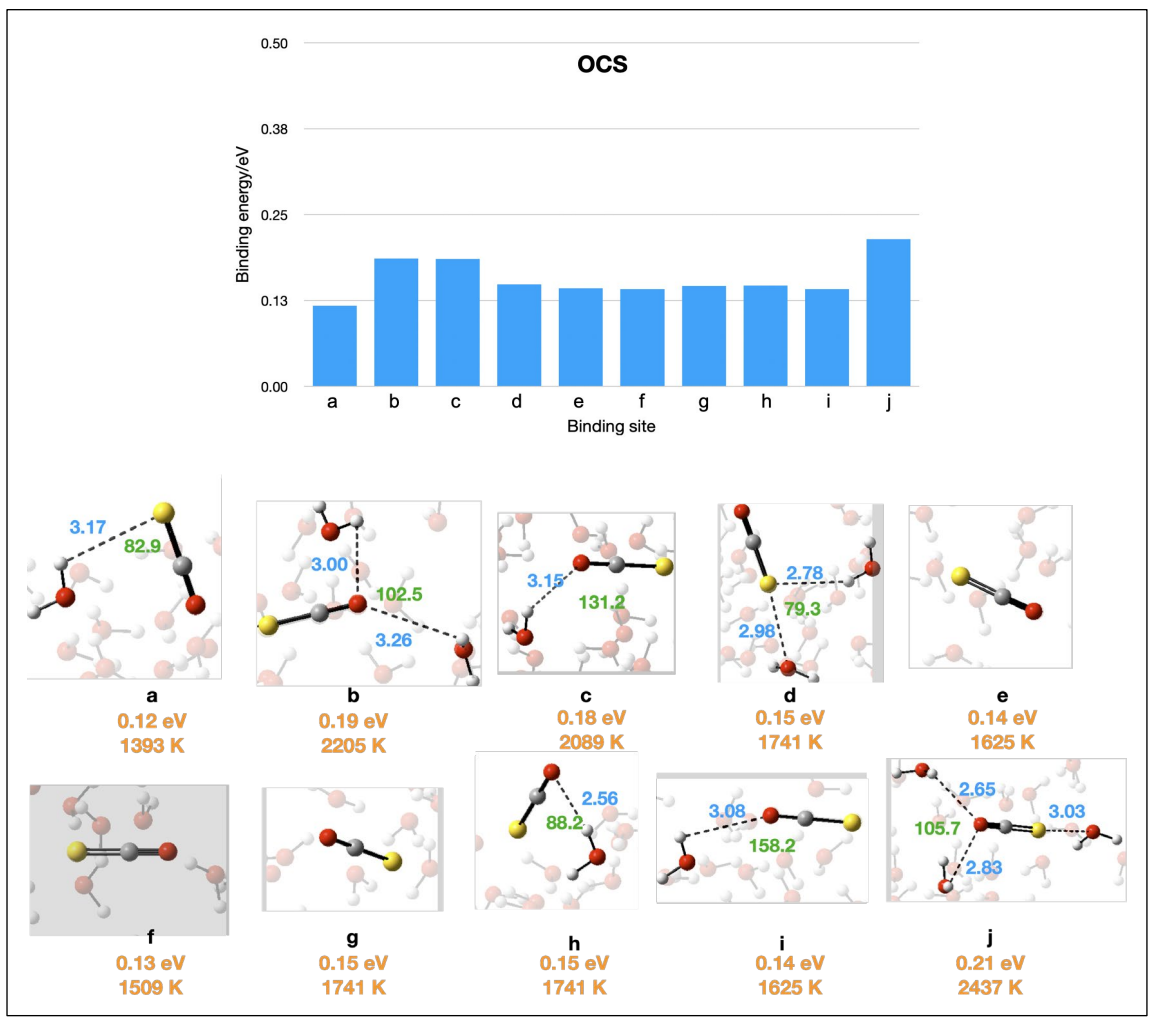

Figure A2. Binding energies of OCS on ASW from wB97X-D/def2-TZVP calculations and optimized structures of OCS on the absorption sites. Calculated binding energies are in eV and K. Only the binding pocket is shown for simplicity. Bond lengths are in Å and bond angles are in degrees. In general, binding energy becomes stronger when OCS interacts with the dangling-Hs on ice (e.g., binding sites b and j). When OCS does not interact with the dangling-H on ice, for instance binding sites e, f, ang g, computed binding energy is weak.

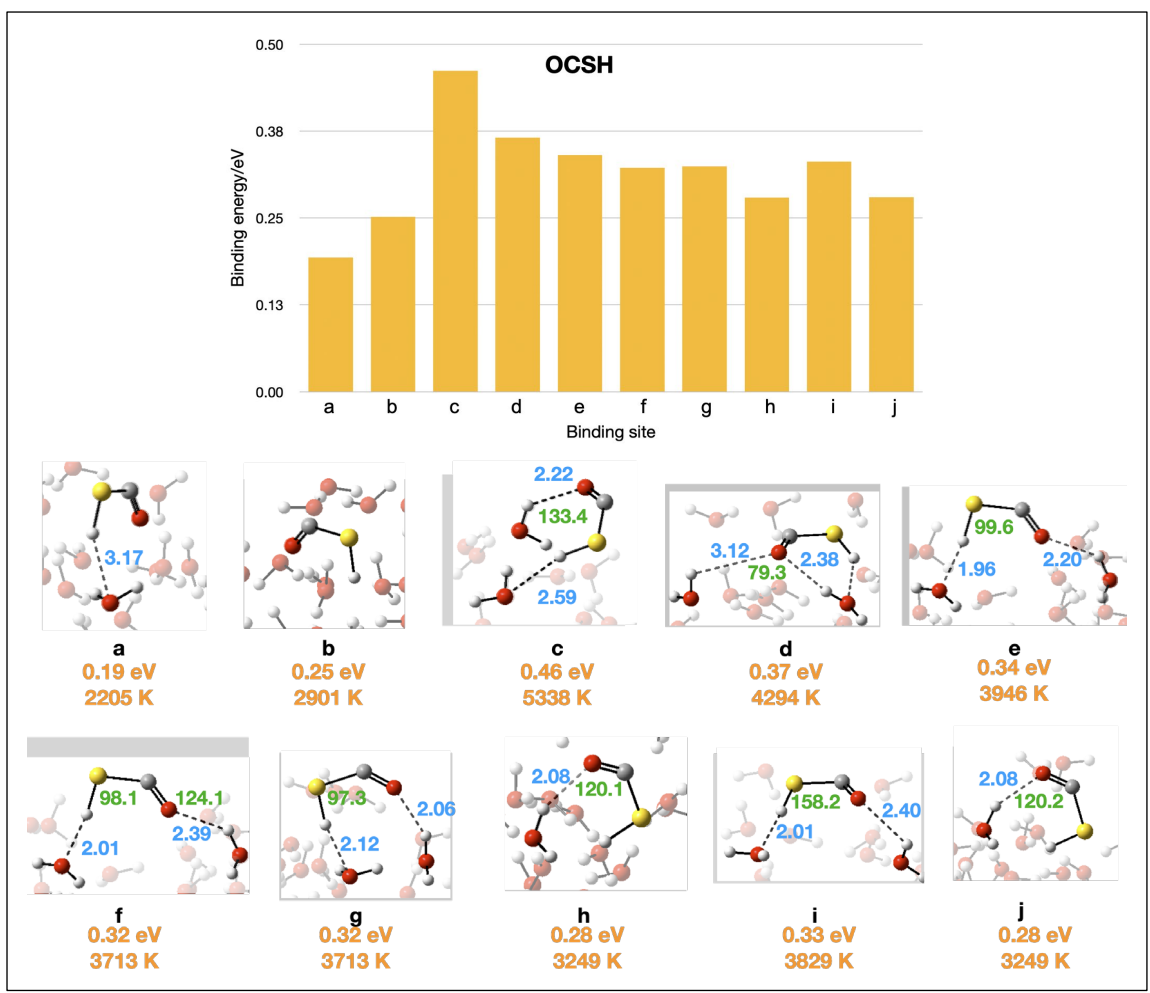

Figure A3. Binding energies of OCSH on ASW from wB97X-D/def2-TZVP calculations and optimized structures of OCSH on the absorption sites. Calculated binding energies are in eV and K. Only the binding pocket is shown for simplicity. Bond lengths are in Å and bond angles are in degrees. Computed binding energy becomes stronger if the OCSH radical interacts with the both dangling-H and dangling-O on ice (e.g. binding sites c and d).

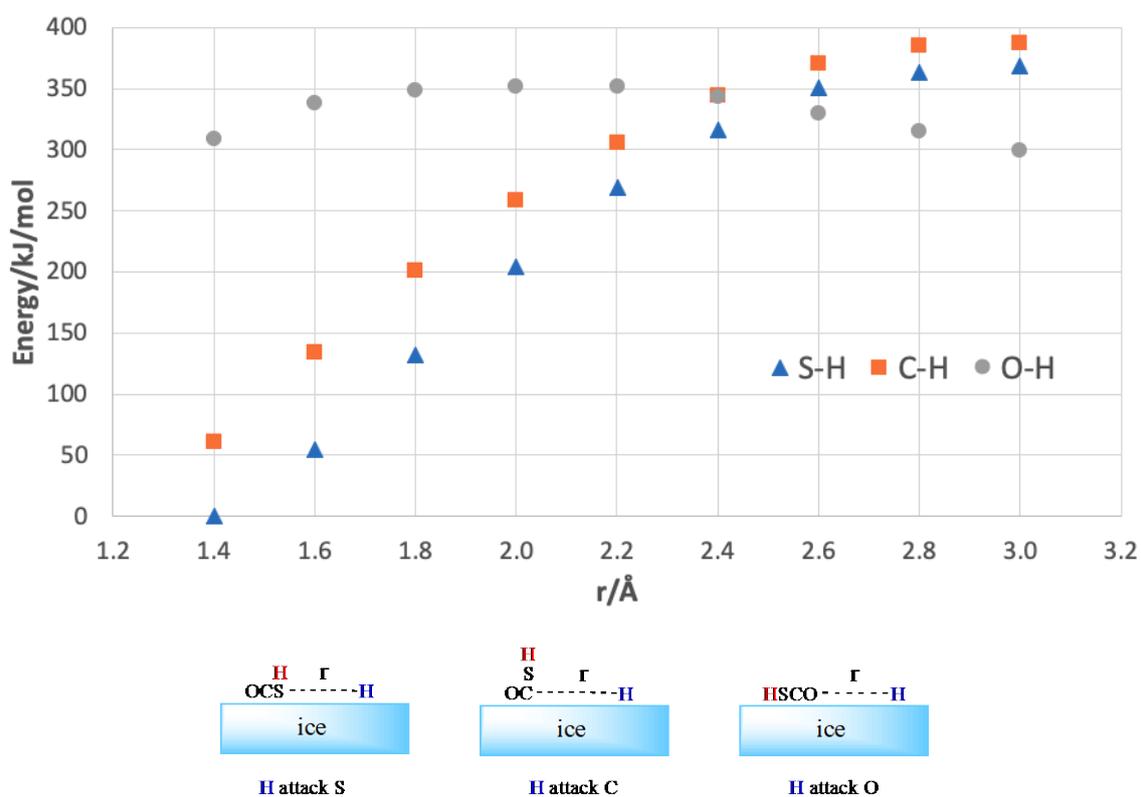

Figure A4. Relaxed potential energy surfaces for the reactions between OCS-H and H from ONIOM(CCSD(t)/aug-cc-pVTZ:wB97X-D/def2-TZVP) calculations, where the two radical species are in the ONIOM high later and the ice structure is in the ONIOM low layer. "r" is the reaction coordinate. At the present level of theory, the relaxed potential energy surfaces for the OCS-H + H → $H_2S$ + CO and OCS-H + H → HOCSH reactions are attractive toward short-range separations. Therefore, these two reactions are barrierless. On the other hand, the relaxed potential energy surface for the OCS-H + H → HOCSH reaction shows a barrier and the product, HOCSH, is relatively unstable compared to $H_2S$ + CO and HOCSH.

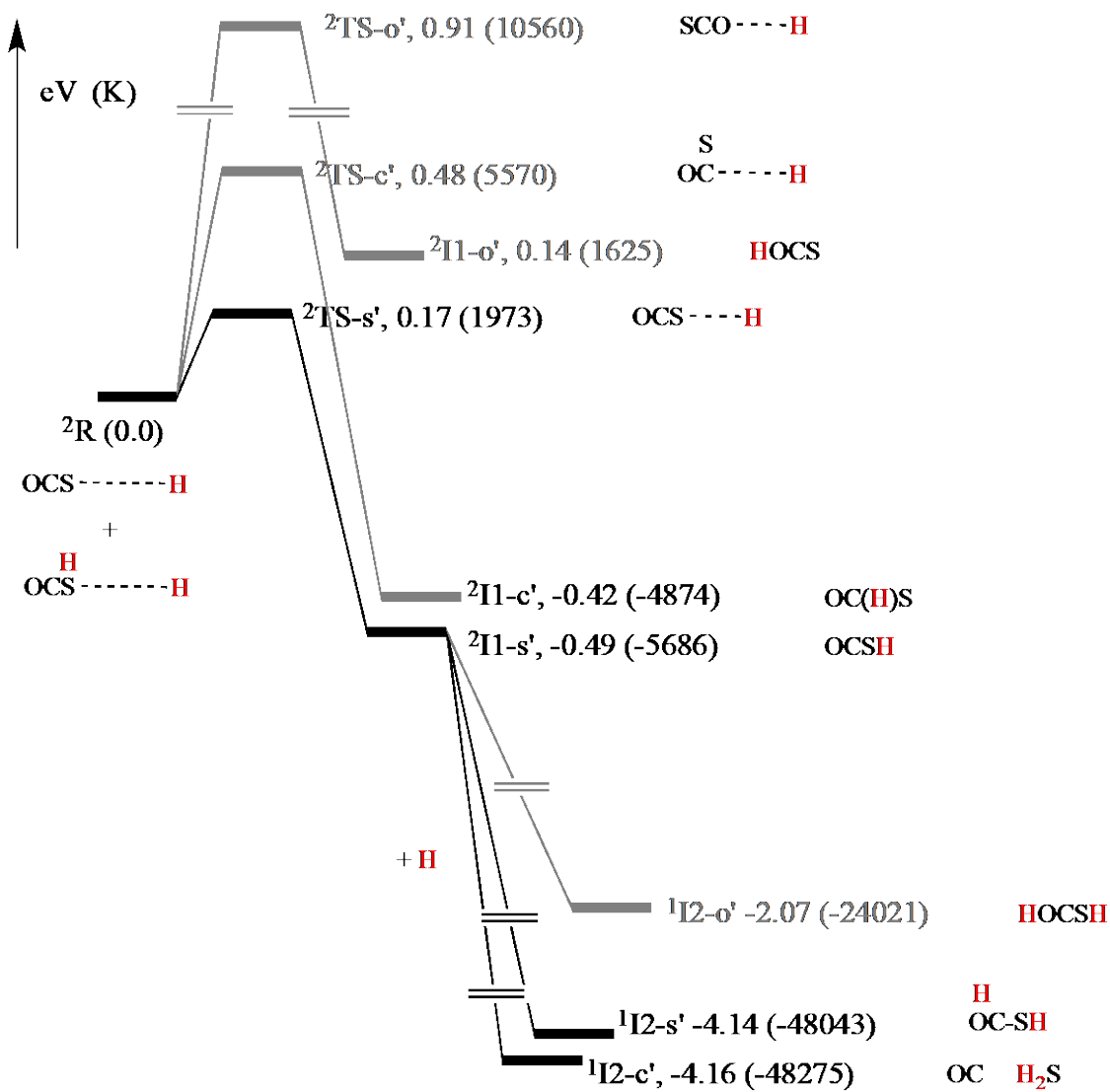

Figure A5. Computed potential energy surfaces for the reaction between OCS and H in the gas phase from wB97X-D/def2-TZVP structure optimizations. The multiplicity of the local minima or transition states is shown in superscript. Potential energy sum of OCS-H (doublet state) and OCS(H)-H (open-shell single state) were taken as the reference energy. Lines in black show the main reaction paths. Reaction between H and OCS would give OCSH radical ($^2$I1-s′), and this reaction has a barrier of 0.17 eV (1973 K). Subsequent reaction paths for the hydrogenation of OCSH radical are highly exothermic, where formation of CO and $H_2S$ would be favorable. The calculated energies of the optimized stationary points and imaginary frequencies are shown in Table A1. Cartesian coordinates of the optimized structures can be found in the tar.gz package of supporting information.

Table A1. Calculated energies of the optimized stationary points and imaginary frequencies from the wB97X-D/def2-TZVP level of theory.

| Stationary point | Potential energy/Hartree | Imaginary frequencies/cm$^{-1}$ |
|---|---|---|
| $^2$R | -5098.452293 | - |
| $^2$TS-s | -5098.444465 | -534.1 |
| $^2$TS-c | -5098.439147 | -985.4 |
| $^2$TS-o | -5098.416435 | -1819.2 |
| $^2$I1-s | -5098.470438 | - |
| $^2$I1-c | -5098.477322 | - |
| $^2$I1-o | -5098.465892 | - |
| H$_2$S-----CO | -5099.118641 | - |
| HC(O)SH | -5099.115619 | - |
| HOCSH | -5099.045761 | - |
| OCS(H)-----H | -5098.97595 | - |
| Gas-phase | | |
| $^2$R | -512.049183 | - |
| $^2$TS-s | -512.042770 | -723.7 |
| $^2$TS-c | -512.031573 | -1025.9 |
| $^2$TS-o | -512.015498 | -1741.0 |
| $^2$I1-s | -512.067260 | - |
| $^2$I1-c | -512.064799 | - |
| ²I1-o | -512.043958 | - |
| $H_2S$-----CO | -512.705283 | - |
| HC(O)SH | -512.705877 | - |
| HOCSH | -512.628452 | - |
| OCS(H)-----H | -512.569855 | - |